\documentclass[]{spie}

\title{Remarks on "Piezonuclear neutrons from fracturing of inert solids"}
\author{G. Comoretto\supit{a}, M. Prevedelli\supit{b}
\skiplinehalf
\supit{a}INAF Osservatorio di Arcetri, Largo E. Fermi 5, Firenze, Italy; \\
\supit{b}
Universit\`a di Bologna, 
Dip. di Fisica e Astronomia,
V. Irnerio 46, Bologna, Italy;
}

\begin{document}
\maketitle

\abstract{
In two series of measurements, Cardone, Carpinteri {\em et al.} report an excess of
neutrons over the background flux corresponding to the 
catastrophic fracture of a granite block subject to compression. Here we
show that these measurements contain large inconsistencies with respect to
the stated experimental procedure, including fractional neutron counts
and strongly non Poissonian statistics. }

\section{Introduction}
In Ref. \citenum{carp_strain, carp_phrev} 
an excess of neutrons over the background flux, corresponding to the 
catastrophic fracture of a granite block subject to compression, is
reported. 

The neutron flux is measured using a single He$^3$ detector, whose count
rates (cps) are reported as graphs versus time. Counts are integrated over 
60 s periods, for the duration of each measurement (approx. 35 minutes).

Another set of measurements, involving 7 granite blocks, with the same
experimental set-up, is reported in Ref. \citenum{carp_physme}. 5 of the
7 measurements show an excess of neutron counts during granite fracture,
but only the 3 most statistically significant, labelled P6, P8 and P9,
are reported in detail.

Both sets show evident anomalies, both in the reported data points and
in their statistics.

\section{Fractional counts}

The neutron flux is reported articles in graph format, in units
of counts per second (cps). These values can be converted to counts
per measurement period (counts per minute, or cpm), with an accuracy of
better than 0.2 cpm. For comparison, the error bars shown in the graph
are at $\pm 0.12$ cpm. As counts are discrete events, the measured values
should be always multiple of 1 cpm, or 1/60 cps.  With a vertical scale
of roughly 1.85 mm/cpm, a quantization in the vertical scale for the
measurement result should be easily appreciable.  On the contrary, all
plots in Ref. \citenum{carp_phrev} show fluctuations that are almost
continuous, up to the graphical resolution of the plots. For example the
background for specimen P3, after the first 5 minutes, undulate almost
continuously between 1.8 and 2.7 counts per integration period.

The flux observed during the fracture of block P4 is reported as 0.272
cps, consistent with the position of the data point in the graph. This
corresponds to 16.3 counts in the measurement period.

Possible explanations for fractional counts could involve a large
number of detectors, but it is explicitly stated that only one 
He$^3$ tube was available.

In Ref. \citenum{carp_physme} the 
background measurements are reported as multiples of $1.6\times
10^{-2}$ counts per second, i.e. integer counts over the 60 seconds
acquisition time. But while for P6 measurements are reported every 60~s,
for the shorter measurements P8 and P9 they are reported every 15~s. This
is inconsistent with the experiment description, but if the acquisition
time was actually shortened, the reported values would result
in fractional counts. Measures do not overlap, as the excess neutron
flux can be seen in just one data point. 

\section{Statistical analysis}

In Ref. \citenum{carp_phrev} the background neutron flux is reported as
$3.8\pm0.2\times10^{-2}$ cps. The stated uncertainty is 4 times lower than
the expected standard deviation for a Poissonian distribution with 23
measured neutrons, as can be derived from the integration time of 600 s.
This uncertainty is used for all measurements in the article, despite
the still much lower statistics for the background and event samples,
that should give statistical errors at least 12 times larger.

The scatter in the background flux is incredibly low. A Poissonian
statistics with a mean of 2.3 events per period should involve a much
larger scatter. In a 35 minute period, one should observe a few periods
with no counts at all, and a few with 5 or more counts ($7.3\times10^{-2}$
cps), while no such occurrences are visible in the four graphs shown. The
variance in the background measurements for specimen P3, for example,
is around 0.25 (cpm)$^2$, 8 times less than the expected value
for Poissonian events. 
Fluctuations are much lower, usually less than 0.2 counts, over most of
the background graphs for all specimen. The time series, in all graphs, 
show also a very strong correlation in time, up to several minutes,
not consistent with background random radioactive events. 

Analysis of Ref. \citenum{carp_physme} depends on which of the assumptions
on the measurement time are correct. Assuming a measurement time of 15
s would be absolutely inconsistent with a Poissonian statistics, so 60 s
has been assumed. In this case measurement statistics is comparable among
samples, consistent with the almost uniform background flux measured.
The statistics of all the 77 background data points
plotted in the three graphs has thus been carried together, as a single
population.
From these measures, a background flux of $4.0\pm0.45\times10^{-2}$
cps can be derived, again consistent with the background fluxes listed
in their tab. 3. For subsequent analysis, a background flux of
$4.2\times10^{-2}$ cps has been assumed. 

The distribution is however quite odd: it is reasonably consistent
with a Poisson distribution, for both the above values of the background
flux, except that there are no occurrences of
measurement periods with zero detections. The expected number
of such occurrences would be between 6.2 and 7, among the 77 background
data points plotted in Ref. \citenum{carp_physme}, with a probability of
less than 0.2\% of observing no such occurrences.

\section{Conclusions}

The data presented in the analyzed articles by Carpinteri
{\em et al.} cannot be the result of a measurement using a counting
detector. The first set of measurements shows non-integer counts in the
reported measurements. The figures for samples P8 and P9 of
Ref. \citenum{carp_physme} show 4 times more points than those actually
measured. The measurement statistics is completely meaningless in the
first set of measurements, and inconsistent with a Poisson statistics
in the second.

\bibliography{carp}

\begin{thebibliography}{1}

\bibitem{carp_strain}
Carpinteri, A., Cardone, F., and Lacidogna, G., ``Piezonuclear neutrons from
  brittle fracture: Early results of mechanical compression tests,'' {\em
  Strain}~{\bf 45},  332--339 (2009).

\bibitem{carp_phrev}
Cardone, F., Carpinteri, A., and Lacidogna, G., ``Piezonuclear neutrons from
  fracturing of inert solids,'' {\em Physics Letters A}~{\bf 373},  4158--4163
  (2009).

\bibitem{carp_physme}
Carpinteri, A., Borla, O., Lacidogna, G., and Manuello, A., ``Neutron emissions
  in brittle rocks during compression tests: Monotonic vs. cyclic loading,''
  {\em Physical Mesomechanics}~{\bf 13.5-6},  268--274 (2010).

\end{thebibliography}
\bibliographystyle{spiebib}   

\end{document}